**Ultrathin Magnesium-based Coating as an Efficient Oxygen Barrier for Superconducting Circuit Materials**

Chenyu Zhou[1], Junsik Mun[1,2], Juntao Yao[2,3], Aswin kumar Anbalagan[4], Mohammad D. Hossain[5], Russell A. McLellan[6], Ruoshui Li[1,7], Kim Kisslinger[1], Gengnan Li[1], Xiao Tong[1], Ashley R. Head[1], Conan Weiland[8], Steven L. Hulbert[4], Andrew L. Walter[4], Qiang Li[2,9], Yimei Zhu[2], Peter V. Sushko[5], Mingzhao Liu[1]

[1]Center for Functional Nanomaterials, Brookhaven National Laboratory, Upton, New York 11973

[2]The Condensed Matter Physics and Materials Science Department, Brookhaven National Laboratory, Upton, New York 11973

[3]Department of Materials Science and Chemical Engineering, Stony Brook University, Stony Brook, New York 11794, United States

[4]National Synchrotron Light Source II, Brookhaven National Laboratory, Upton, New York 11973

[5]Physical and Computational Sciences Directorate, Pacific Northwest National Laboratory, Richland, Washington 99354

[6]Department of Electrical and Computer Engineering, Princeton University, Princeton, New Jersey 08540

[7]Department of Chemistry, Stony Brook University, Stony Brook, New York 11794

[8]Materials Measurement Laboratory, National Institute of Standard and Technology, Gaithersburg, Maryland 20899

[9]Department of Physics and Astronomy, Stony Brook University, Stony Brook, New York 11794

*Correspondence: zhu@bnl.gov (Y.Z.), peter.sushko@pnnl.gov (P.V.S.), and mzliu@bnl.gov (M.L.)

**Abstract**

Scaling up superconducting quantum circuits based on transmon qubits necessitates substantial enhancements in qubit coherence time. Among the materials considered for transmon qubits, tantalum (Ta) has emerged as a promising candidate, surpassing conventional counterparts in terms of coherence time. However, the presence of an amorphous surface Ta oxide layer introduces dielectric loss, ultimately placing a limit on the coherence time. In this study, we present a novel approach for suppressing the formation of tantalum oxide using an ultrathin magnesium (Mg) capping layer deposited on top of tantalum. Synchrotron-based X-ray photoelectron spectroscopy (XPS) studies demonstrate that oxide is confined to an extremely thin region directly beneath the Mg/Ta interface. Additionally, we demonstrate that the superconducting properties of thin Ta films are improved following the Mg capping, exhibiting sharper and higher-temperature transitions to superconductive and magnetically ordered states. Based on the experimental data and computational modeling, we establish an atomic-scale mechanistic understanding of the role of the capping layer in protecting Ta from oxidation. This work provides valuable insights into the formation mechanism and functionality of surface tantalum oxide, as well as a new materials design principle with the potential to reduce dielectric loss in superconducting quantum materials. Ultimately, our findings pave the way for the realization of large-scale, high-performance quantum computing systems.

**Introduction**

The most recent two decades have witnessed rapid development of infrastructure for quantum computation driven by the advances of superconducting quantum circuits (SQC).[1,2] The core element in a SQC is the superconducting qubit that centers around one or more Josephson junctions and acts as a quantum two-state system.[3,4] Among a variety of superconducting qubit designs, the *transmon* architecture, in which the Josephson junction is shunted by a large capacitor for charge noise suppression, is one of the most promising platforms for realizing large-scale quantum computations.[5-7] For the purpose of scalability, achieving long coherence time is always the priority in the research on transmon qubits. With this goal, a number of superconducting materials have been tested as constituents of transmon qubits, represented by elementary metals including aluminum (Al)[8,9], niobium (Nb)[10], and recently reported tantalum (Ta)[11,12]. Although the synthesis process of superconducting thin film is generally performed at high vacuum, the oxidation of metal surface is inevitable upon exposure to the atmosphere. The surface oxide layer, typically disordered, is thought to contain defects that exhibit two-level system (TLS) behavior, which leads to dielectric loss and thus decreases qubit coherence time.[13,14] Therefore, it is of importance to obtain insights into the atomic-scale mechanisms that lead to the formation of the surface oxide layer in the superconducting materials and identify approaches for suppressing its growth.

It was recently shown that body-centered-cubic (*bcc*) tantalum (α-Ta), a type-I superconductor with $T_c$ = 4.0 K to 4.4 K[15], is a promising material to improve the coherence time of transmon qubit.[11,12] Place et al. demonstrated that, by replacing the Nb capacitor and resonator with α-Ta, the $T_1$ lifetime of transmon qubit can be extended beyond 300 μs.[11] Through design and fabrication improvements involving a dry-etching process during microfabrication, Wang et al. further increased the $T_1$ lifetime of tantalum transmon qubit to > 500 μs.[12] A primary advantage of Ta over Nb is attributed to the formation of a self-limiting native oxide in the former that is thinner and dominated by the fully oxidized, insulating $Ta_2O_5$, which reduces dielectric loss at the metal-air interface.[4,11] In contrast, Nb forms a thicker native oxide layer that is chemically more complicated, thereby potentially providing more loss channels.[10,16-18] In a recent study, Bal et al. passivated the Nb surface with Ta, which led to a significant improvement in $T_1$ coherence time.[19] Their findings provide direct evidence verifying the advantage of the native oxide of Ta over that of Nb in reducing dielectric loss. Even though tantalum oxide provides better performance than niobium oxide, one cannot ignore uncontrolled defects and other likely loss channels in the oxide layer.[20] Therefore, it is important to establish approaches to suppress the formation of surface oxide on tantalum thin film. One feasible approach is coating Ta with another material to protect the tantalum metal from oxidation.

An ideal coating material should be stable in both deposited and oxidized forms, immiscible with tantalum, and form a barrier to oxygen diffusion into the Ta film. Magnesium (Mg) is a good candidate that satisfies these criteria. First, according to the equilibrium phase diagram of Mg and Ta, it is quite unlikely to form Mg/Ta alloy (Figure S1).[21] Second, the growth of magnesium is facile in most deposition tools.[22,23] Third, the oxidation reaction between magnesium oxide and metal tantalum is thermodynamically unfavorable.[21,24] In this work, we *in vacuo* deposit a tantalum thin film and then a magnesium capping layer using magnetron sputtering, avoiding the exposure of the fresh tantalum surface to air. We demonstrate that a very thin layer of magnesium (3 nm) coating can effectively suppress the formation of $Ta_2O_5$ even though the Mg layer is fully oxidized to MgO in air. Specifically, detailed analysis of the MgO/Ta interface using electron microscopy and X-ray spectroscopies demonstrates that the magnesium-based coating remarkably reduces the thickness of the tantalum surface oxide species. Our *ab initio* modeling suggests that the magnesium capping layer acts as an oxygen scavenger that facilitates extraction of inadvertent oxygen impurities from the Ta layer and adsorbs environmental oxygen, ultimately forming a coating impermeable to oxygen. As a result, the formation of the tantalum oxide is suppressed, and crystallinity of the tantalum film is improved, as indicated by higher superconducting transition temperature and sharper transition to the superconducting state. Furthermore, based on the Mg-coated tantalum film, some new interfaces are artificially created leveraging the chemical reactivity of magnesium oxide. For example, given the fact that magnesium oxide can be readily removed chemically, we can regenerate Ta oxide at more strictly controlled conditions for more reproducible device fabrication. This work not only offers a novel strategy to protect tantalum from oxidation, but also provides many insights in understanding the formation mechanism and functions of surface tantalum oxide.

**Results and Discussion**

Ta thin films are deposited by magnetron sputtering to a thickness of approximately 150 nm, over A-plane-cut ($11\bar{2}0$) sapphire substrates held at 750 °C. Then the as-grown Ta film is cooled down to about 100 °C at the end of its deposition stage, after which a very thin layer (3 nm) of Mg metal is deposited *in vacuo* over its surface to protect it from oxidation in ambient conditions (Figure 1a). Finally, the sample is cooled to room temperature, removed from the deposition chamber, and exposed to air. Hereafter, the Ta films with and without Mg coating will be denoted as *Native Ta* and *Mg/Ta*, respectively. As revealed by X-ray diffraction (XRD) measurements, the Ta film is grown in pure *bcc* α-phase with (110) out-of-plane orientation, while the Mg coating is too thin to produce distinguishable diffraction signals (Figure S2).

Lab-based X-ray photoelectron spectroscopy (XPS) reveals stark difference between the surface chemistries of Native Ta and Mg/Ta (Figure 1B). In the Ta 4f region, the spectrum of Native Ta can be well fitted with three sets of spin-orbit split doublet peaks ($4f_{5/2}$ and $4f_{7/2}$), using the Gaussian-Lorentzian product line shape. The doublet with the lowest binding energies (B.E. = 21 – 23 eV) is associated with tantalum metal ($Ta^0$), while the one with the highest binding energies (26 – 29 eV) is contributed by $Ta_2O_5$ ($Ta^{5+}$). In addition, there is a smaller doublet located between $Ta^0$ and $Ta^{5+}$ peaks, which may be assigned to Ta suboxides ($Ta^{x+}$) based on its binding energies.[25,26] Similarly to the XPS spectra published elsewhere for native oxide-covered Ta films[11], the $Ta^{5+}$ peaks are much larger than the $Ta^0$ peaks, suggesting the formation of a native oxide of few nanometer thickness. On the other hand, Ta 4f spectrum of Mg/Ta is dominated by the $Ta^0$ peaks, followed by the Ta suboxide peaks, with minimal presence of the $Ta^{5+}$ peaks, indicating that the surface oxidation of Ta is remarkably suppressed by the magnesium-based capping layer (Figure 1B). The XPS signal from Ta remains strong, suggesting that the Mg layer is very thin and contributes little to photoelectron attenuation.

As revealed by transmission electron microscopy (TEM) analysis, while a thin layer (thickness 3.9 nm ± 0.3 nm) of amorphous oxide covers the Native Ta surface (Figure 1C), it is notably absent from the Mg/Ta surface (Figure 1D and Figure S3). In Mg/Ta, the presence of a very thin Mg capping layer is verified by combining the high angle annular dark field – scanning TEM (HAADF-STEM) with core-level electron energy loss spectroscopy (EELS) line scans (Figure 2A and Figure S4). A finer EELS mapping with scan steps of 0.05 nm near the surface region determines that the Mg layer has a thickness 3.0 nm ± 0.1 nm (Figure 2B). In the mapping, the top surface of the Mg layer is marked by the transition from a Pt M-edge EELS feature to an Mg K-edge (amorphous Pt is deposited as a protection layer for focused ion beam sample preparation), while the bottom surface is marked by the transition from the Mg K-edge to a Ta M-edge. When a thicker layer of Mg (thickness 23.4 nm ± 1.9 nm) is deposited atop the Ta film, the Mg film breaks into an island-like morphology, while the bottom region (thickness 1.4 nm ± 0.1 nm) remains contiguous (Figure 2C).

To obtain the atomic scale insight into the structure and stability of Mg films deposited on the Ta (110) surface prior to their exposure to the ambient conditions, we turn to *ab initio* simulations. Shown in Figure 3 are the calculated thickness dependence of the Mg film structure, the corresponding adsorption energies, and the character of charge redistribution. We find that the first Mg plane forms a stable pseudomorphic layer with an energy gain towards a stable state relative to the bulk Mg of approximately 0.5 eV per atom (Figure 3A and 3D). This interaction is accompanied by an electron transfer from the Mg plane into the Ta layer (Figure 3E), as expected based on the bulk photoelectric work functions of Mg (3.68 eV)[27] and Ta (4.16 eV)[28]. Adsorption of the second Mg plane is isoenergetic to the formation of the Mg bulk, while the third and fourth planes are metastable (Figure 3B and 3D), which is attributed to accumulation of the lattice strain. This result is consistent with the experimental observation on a thicker Mg capping layer, which remains contiguous at its bottom, but breaks into islands beyond a height of 1.4 nm. Analysis of the angles between Mg-Mg bonds (Figure 3C) shows that the Mg film reconstructs from the *bcc* lattice near the interface to the *hcp* lattice, as expected for the bulk Mg. Interestingly, while the amount of charge transferred from Mg into the Ta layer increases with Mg coverage up to 3 monolayers (ML, Figure 3E), most of this charge is associated with the first interfacial Mg plane. Together with only a slight corrugation of the interfacial Mg plane with increasing Mg thickness (Figure 3A and 3B) and its relatively high stability, we conclude that the key characteristics of the Mg/Ta (110) interface are well captured by the single Mg plane. Our results also suggest that even if thicker Mg films become amorphous to accommodate the lattice mismatch strain, the interfacial Mg plane is likely to remain ordered.

The chemical profile of Ta species across the surface region of Mg/Ta is further quantified using a novel strategy based on synchrotron variable-energy XPS (VEXPS) recently developed by us.[26] In this technique, a material surface is studied by XPS using a wide range of incident photon energies. The surface probing depth is thus tuned by varying the photon energy of the incident X-ray beam. At the higher photon energy end, photoelectrons are generated with higher kinetic energy and longer mean free paths, thus giving XPS more bulk sensitivity. *Vice versa*, XPS at the lower photon energy end has more surface sensitivity. In this work, we use 15 photon energies between 760 eV and 5000 eV to probe the Ta 4f spectral region. At each incident photon energy, the photoelectron intensity fractions of each Ta species, either oxide or metal, are derived through peak fitting (Figure 4A-C and Figure S5). The incident energy dependence of Ta 4f spectra is largely captured by fitting with concentration profiles of three unique Ta species, namely $Ta^{5+}$ (fully oxidized Ta), $Ta^{x+}$ (Ta suboxides), and $Ta^0$ (metallic Ta). Following the technique we previously developed, the concentration profiles are constructed using smooth analytical functions as fitting basis (Figure 4D).[26] According to the fitted concentration profile, the Mg coating reduces the integrated thickness for Ta surface species when compared with Native Ta. Ta oxides in Mg/Ta are limited to an ultrathin region just beneath the Mg/Ta interface (Figure 4E), amounting to a total integrated thickness below 1 nm, which contrasts strongly against Native Ta that has an oxide layer over 2 nm in the integrated thickness (Figure 4F). Interestingly, the Ta oxide in Mg/Ta is dominated by $Ta^{x+}$ rather than $Ta^{5+}$, suggesting a lowered oxygen chemical potential. Overall, it can be concluded that just a 3-nm thick Mg layer is able to greatly suppress the oxidation of tantalum.

As expected, the thin Mg layer is eventually oxidized after exposure to air. The binding energy of Mg 2p locates at 50.8 eV and agrees well with the literature value of MgO (Figure 5A).[29-31] Two unique components are resolved from the O 1s spectrum, with a minor one located at 530.5 eV and a major one at 532.6 eV (Figure S6), which correspond to lattice $O^{2-}$ and hydroxyl groups, respectively. This finding suggests that the air exposure not only oxidizes Mg but also hydroxylates it.[32] The distribution of hydroxyl groups in the surface region is investigated by analyzing the O 1s spectral region using VEXPS at various incident photon energies (Figure S7). As the incident photon energy decreases, the XPS peak area fraction of the hydroxyl group OH increases, indicating that hydroxyl groups are located closer to the top surface (Figure S8). Our calculations confirm that the Mg layer in Mg/Ta (110) system is readily oxidized upon exposure to an $O^{2-}$-containing atmosphere. To understand the atomic-scale processes responsible for this behavior, we examine changes of the oxygen binding energies and charge redistribution with increasing oxygen content. To this end, we define incremental binding energies ($E_i$) that reflect energy gain due to incorporation of each individual O atom for a given oxygen coverage. The $E_i$ values are sensitive to the local structure of the oxygen binding site. For the oxidation of a 1 ML (defined by the number of Ta atoms per lateral supercell) Mg film, $E_i$ fluctuates between approximately 3 eV and 5.5 eV as the O coverage increases from zero to 2 ML (Figure 5B). To reflect the overall energy gain due to oxidation, we further define average binding energy $E_a(\sigma) = \sigma^{-1} \int_0^\sigma E_i(x)dx$ for each O coverage ($\sigma$), which varies much less (4 eV to 4.5 eV) across the same O coverage window. The calculation shows that $E_a$ increases as O coverage increases to 1 ML, which formally corresponds to stoichiometric MgO (Figure 5B). In other words, the emerging electrostatic interactions in the $MgO_x$ layer promote oxygen trapping. However, the trend is reversed as O coverage increases further. These changes can be rationalized in terms of charge density redistribution as represented by the total charges of the Ta and Mg regions (Figure 5C). We identify three stages of the oxidation process, labeled as I, II, and III in Figure 5C and 5D. At stage I, positive charge of the Mg species increases with increasing O content, which indicates that charge density redistribution proceeds within the $MgO_x$ layer only. At stage II, Mg species approach their fully oxidized state as O content approaches 1 ML, indicating the MgO layer formation. In addition, electron charge that was transferred to Ta upon Mg/Ta (110) formation is pulled from the Ta layer back into the film, rendering the charge of Ta layer neutral (Figure 3E). Finally, further binding of the oxygen atoms is facilitated by the electron transfer from the outermost plane of the Ta layer only (stage III). As the positive charge of this plane increases, electron transfer to the additional O becomes less thermodynamically favorable, which eventually suppresses the oxidation process. Schematics of this process is shown in Figure 5D. In summary, in the case of 1 ML Mg film, 2 ML O coverage results in the formation of an amorphous MgO layer and a nearly ordered $TaO_x$ layer in the outer region of the Ta layer.

After exposure to air for 30 days, the lab-based XPS of Mg/Ta remains very similar as freshly deposited sample. Angle-resolved XPS (AR-XPS) is conducted to study the degree of oxidation of the surface region of Mg/Ta, with enhanced surface sensitivity when decreasing the collection angle of photoelectron by tilting sample stage (Figure S9). Only through AR-XPS at high tilting angle, which is most surface-sensitive, we could tell that the aged sample contains slightly more $Ta^{5+}$ (Figure 6A and Figure S10). Similarly, lab-based XPS observes no sign of Ta oxidation after the Mg/Ta film is annealed in air for 5 minutes at 200 °C (Figure S11). The temperature threshold for oxygen breakthrough is determined by ambient-pressure XPS (AP-XPS), in which XPS spectra are collected *in situ* from a Mg/Ta thin film gradually heated in an ambient of oxygen at 50 Pa. The oxidation onset of Ta is observed at 300 °C, a point that coincides with the complete strip-off of surface hydroxyl groups (Figure S12). We also note that Ta is not oxidized during high vacuum annealing without $O_2$ present, which rules out the possibility of the reaction between MgO and Ta (Figure S13). Therefore, it can be concluded that the MgO layer serves as an effective kinetic barrier in suppressing Ta oxidation.

First-principles calculations suggest that the Mg/MgO$_x$ coatings provide a dual protection of the Ta substrate, with the primary role as an effective barrier preventing O penetration from an O-rich atmosphere. To illustrate this effect, we calculate the potential energy profiles for an $O^{2-}$ ion transferred from the MgO$_x$ layer (1 ML Mg) into Ta for several O coverage values (Figure 6B). When the O coverage is below 2 ML, oxygen binds to the surface with the energy gain of 3 – 4 eV relative to the gas-phase $O_2$ and adopts the lowest energy configurations at the MgO$_x$/Ta interface. The transfer of this $O^{2-}$ ion into Ta substrate is endothermic by 0.7 – 1 eV depending on the MgO$_x$ oxidation state and is therefore energetically unfavorable. In contrast, when the O coverage exceeds 2 ML, dissociative adsorption of $O_2$ on the MgO$_x$ surface is endothermic by as much as 0.6 eV (Figure 6B). Should the dissociative adsorption occur, interstitial O ($O_i$) that binds to pre-existing $O^{2-}$ will emerge in the form of peroxide ions $O_2^{2-}$. According to our calculations, such interstitial readily diffuses towards the MgO$_x$/Ta interface and converts to an $O^{2-}$ ion by trapping electrons from the Ta layer. We find that the most stable location for this $O_i$ is in the Ta subsurface, i.e., it initiates oxidation of another Ta atomic plane, which is similar as the pure Ta oxidation process discussed elsewhere.[33] Therefore, although Ta oxidation becomes energetically favorable after the complete oxidation of Mg and the formation of MgO/TaO$_x$ interface, the 0.6 eV energy barrier for $O_2$ splitting on the surface of fully oxidized MgO effectively suppresses O transfer into the underlying Ta.

In addition, we identify a secondary role of Mg as a getter for residual oxygen trapped in Ta. Experimentally, it is observed that when a thick (~20 nm nominally) Mg layer is deposited over Ta, so that the top surface and the bottom surface of Mg can be clearly differentiated by TEM, the EELS area mapping reveals oxidation of Mg at the Mg/Ta interface while the underlying Ta remains oxygen-free (Figure 2D). Since the air exposure only oxidizes the top 2 – 3 nm of Mg, the oxidation of Mg at the Mg/Ta interface can only be explained by its gettering action on Ta, which may have trace amount of O trapped during sputtering deposition, due to the imperfect vacuum (base pressure typically about $10^{-6}$ Pa). The calculated potential energy profiles for the transfer of residual oxygen ($O_i$) from the Ta substrate to the surface or the Mg/Ta interface (Mg coverage is 0 ML to 2 ML) are shown in Figure 6C. The calculated energies are aligned to set the energy of the $O_i$ in the inner part of the Ta slab to zero. From these calculations, we conclude that residual O in Ta is stabilized by as much as 1.5 eV if it is transferred to the surface. Hence, residual O species are expected to be expelled from the bulk Ta during Mg deposition and be stabilized in the Mg film.

Low-temperature transport measurements indicate that the Mg capping layer improves the superconducting properties of the tantalum thin film. Compared with Native Ta that has superconducting transition temperature $T_c$ = 4.18 K ± 0.02 K, the superconducting transition of Mg/Ta (4.39 K ± 0.01 K) is both sharper and at higher temperature (Figure 7A, inset). For $T_c$ measurements, the error bar represents the half-width of the temperature transition. Moreover, the residual-resistance ratio (*RRR*) of Mg/Ta, defined as the ratio between the resistance at room temperature and just before the superconducting transition, is 10.7 and higher than the *RRR* of Native Ta (7.5). These improved values are consistent with the gettering effect of Mg that removes oxygen impurity from Ta, as observed by the STEM-EELS studies. In addition, the change in dc electrical properties is echoed by the findings in dc magnetic susceptibility measurements. As shown in Figure 7A, the Mg/Ta system exhibits a sharp transition to the Meissner state, characterized by its perfect diamagnetism, immediately below its $T_c$. On the other hand, the transition of Native Ta to its Meissner state comes with a broad tail, not completing until about 0.6 K below its $T_c$. From the measurements of critical field (Figure 7B and Figure S14a), we find that both films fall in the crossover boundary of type-I and type-II superconductor, with the estimated Ginzburg-Landau (GL) parameter, $\kappa$, at 0.78 for Mg/Ta and 0.97 for Native Ta, respective, slightly higher than $1/\sqrt{2} \approx 0.707$. Detailed estimation can be found in the Supplemental Information. The estimated effective coherence length at 0 K, $\xi(0)$, are 45 nm for Mg/Ta and 36 nm for Native Ta films. Lower value of $\xi(0)$ in the Native Ta film is consistent with higher level of impurity leading to shorter mean free path that also produces higher *RRR*.

The amorphous Ta oxide layer can be regenerated after removing the Mg capping layer, which enables us to form tantalum oxide/tantalum interfaces in a controllable fashion. This is achieved by dipping Mg/Ta in a piranha solution (1:2 mixture of 30% hydrogen peroxide and sulfuric acid) for 30 min at room temperature (Piranha-Mg/Ta). According to the lab-based XPS results, there is no Mg left on Piranha-Mg/Ta (Figure S15a) and the surface is now dominated by $Ta^{5+}$ (Figure S15b), revealing that Ta surface is completely oxidized by piranha. Likewise, the OH components are disappeared from the O 1s spectrum as $Ta_2O_5$ surface is not hydroxylated (Figure S15c). Using the synchrotron VEXPS technique, we reconstruct the chemical profile over depth of Piranha-Mg/Ta (Figure S16). The oxide layer of tantalum has similar chemical composition as Native Ta, but is more than 30% thicker by the integrated effective thicknesses (Figure 4F and Figure S16), likely due to the much stronger oxidation ability of Piranha. Nevertheless, the formation of $Ta_2O_5$ remains self-limiting. Along with the regeneration of Ta oxide, we observe a retrogression of $T_c$ to 4.18 K ± 0.02 K (Figure S17). Meanwhile, the critical field $H_{c2}(0)$ also turns up to 0.30 T (Figure S14b). These results confirm the correlation between superconducting properties and surface oxide.

**Conclusion and Outlook**

We have discovered that a thin capping layer of MgO (3 nm) can significantly suppress the formation of tantalum oxide, which is known to contribute to microwave dielectric losses and thus limit the coherence time of tantalum-based qubits. Through quantitative depth profiling using synchrotron VEXPS, we have observed that the presence of tantalum oxides in Mg/Ta is confined to an extremely thin region directly below the Mg/Ta interface, with a total integrated thickness of less than 1 nm. This is in stark contrast to native tantalum, which typically exhibits an oxide layer with an integrated thickness exceeding 3 nm.[26] Additionally, it is worth noting that Mg capping results in sharper transition in superconductivity and magnetization with higher transition temperatures, suggesting that the superconducting properties of the tantalum thin film are improved after the Mg capping. Furthermore, comprehensive testing has revealed that the temperature threshold for oxidation of this Mg-capped system exceeds 200 °C, suggesting that it can withstand the high-temperature processes involved in qubit fabrication, such as resist baking in lithography. Therefore, the engineering strategy of Mg capping holds promise for reducing the dielectric loss channel in Ta-based qubit device and represents a design principle for complex materials systems whereby high crystallinity and purity of a functional component can be ensured by complementing it with a thin sacrificial layer that serves as a chemically inert and impermeable protector against environmental contaminants. More broadly, by refining our understanding of the underlying mechanisms and pushing the boundaries of materials engineering, this study helps to unlock the full potential of quantum technologies and pave the way for the realization of large-scale, high-performance quantum computing systems.

**ORCID**

Chenyu Zhou: 0000-0002-2749-4739

Aswin kumar Anbalagan: 0000-0001-5511-2083

Juntao Yao: 0000-0002-4339-1027

Ruoshui Li: 0000-0002-7903-7070

Xiao Tong: 0000-0002-5567-9677

Conan Weiland: 0000-0001-6808-1941

Qiang Li: 0000-0002-1230-4832

Yimei Zhu: 0000-0002-1638-7217

Peter V. Sushko: 0000-0001-7338-4146

Mingzhao Liu: 0000-0002-0999-5214

**Acknowledgement**

This project is supported by the U.S. Department of Energy (DOE), Office of Science (SC), National Quantum Information Science Research Centers (NQISRCs), Co-design Center for Quantum Advantage ($C^2QA$) under contract number DE-SC0012704. The resources used in this research included the National Institute of Standards and Technology (NIST) operated beamlines Spectroscopy Soft and Tender Beamlines (SST-1 and SST-2), located at the National Synchrotron Light Source II (NSLS-II) and Materials Synthesis & Characterization and Electron Microscopy facilities of the Center for Functional Nanomaterials (CFN), DOE Office of Science Facilities at Brookhaven National Laboratory (BNL) under contract number DE-SC0012704. The authors acknowledged the use of facilities of Electron Microscopy and Nanostructure Group and Advanced Energy Materials Group (Q.L., J.Y.), Department of Condensed Matter Physics & Materials Science at BNL, primarily supported by the DOE, Office of Basic Energy Science (BES), Materials Science and Engineering Division, under contract number DE-SC0012704. Computational modeling at the Pacific Northwest National Laboratory (PNNL) was supported by $C^2QA$ (BES, PNNL FWP 76274). This research used resources of the National Energy Research Scientific Computing Center (NERSC), a DOE SC User Facility supported by the SC of the U.S. DOE under Contract No. DE-AC02-05CH11231 using NERSC Award No. BES-ERCAP0021800. We also thank Nathalie de Leon for useful discussions.

**Author Contributions**

Conceptualization: M.L.; Materials synthesis: C.Z and R.L.; Electron microscopy experiments: J.M. and K.K.; Low-temperature transport measurements: C.Z. and J.Y.; Magnetization measurements: J.Y. and Q.L.; Theoretical calculations: P.V.S. and M.D.H.; Synchrotron-based XPS experiments: C.Z., A.A, and C.W.; Lab-based XPS experiments: C.Z., G.L., X.T., and A.R.H.; XPS data analysis: C.Z. and R.A.M.; Supervision: M.L., P.V.S., Y.Z., Q.L., A.L.W., and S.L.H.; Manuscript draft: C.Z., P.V.S., and M.L.; All authors participated in editing the manuscript.

**Declaration of Interests**

The authors declare no conflict of interest.

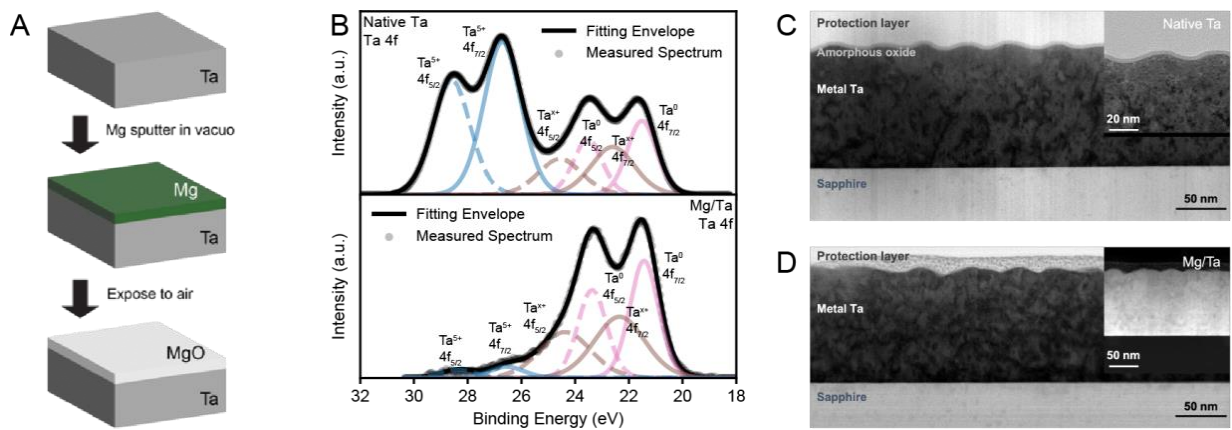

Figure 1. A) Schematic of the *in vacuo* deposition of Mg/Ta thin film. The Mg layer is instantly oxidized upon exposure to air; B) Ta 4f core-level XPS spectra of Native Ta (top) and Mg/Ta (bottom) thin films. Each spectrum is fitted with three sets of spin-orbit split doublet peaks, which are assigned as $Ta^{5+}$ ($Ta_2O_5$), $Ta^{x+}$ (Ta suboxides), and $Ta^0$ (metallic Ta), respectively. The fitting envelope is plotted in solid line, which well overlaps the measured spectrum. The residual standard deviation of fitting is as low as unity; STEM annular bright field (ABF) images of the cross-section of C) Native Ta and D) Mg/Ta thin film. The insets are magnified images of surface regions under TEM and high-angle annular dark-field (HAADF) STEM, respectively. A 3.9 nm ± 0.3 nm thick amorphous layer of tantalum oxide is on the top of Native Ta. The Mg capping layer is too thin to be visible.

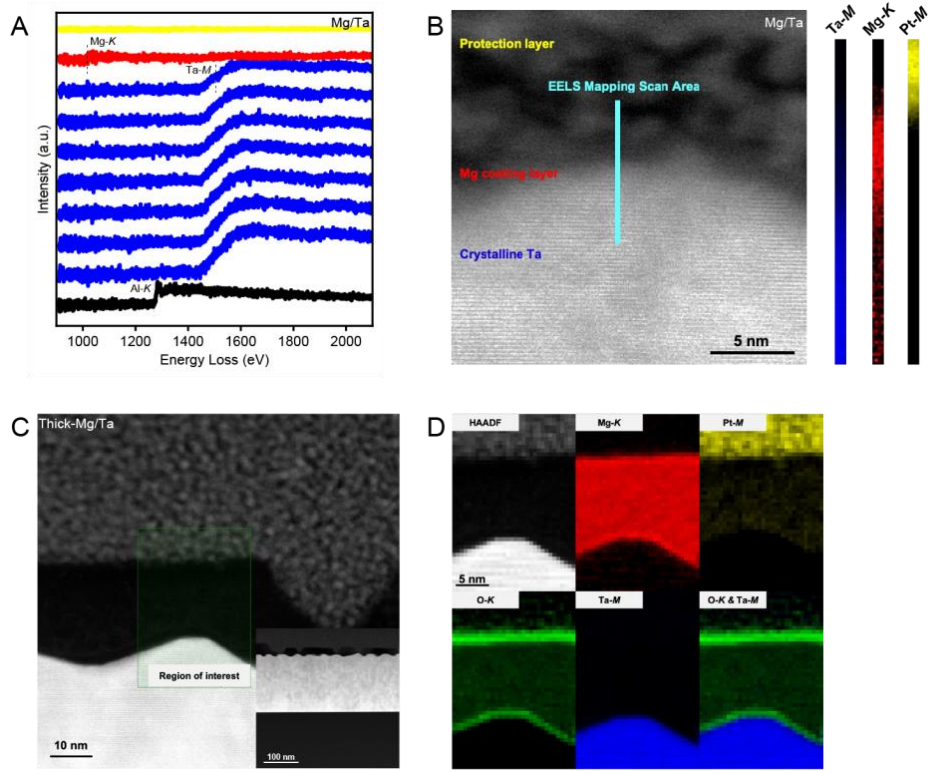

Figure 2. a) EELS line scan spectra of Mg/Ta surface region. The line color corresponds to the number color of the scanning location labelled in Figure S4. Three kinds of features can be resolved from top to bottom: Mg K-edge, Ta M-edge and Al K-edge. The detection of core-level loss spectra at energy range above 1 keV is enabled by K3 detector to overcome the difficulty of high background intensity and low signals of the edges in EELS; b) *Left*: High-resolution cross-sectional HAADF-STEM image of the surface region of Mg/Ta thin film; *Right*: EELS mappings of Ta M-edge, Mg K-edge and Pt M-edge loss. In the HAADF image, the thin rectangle marks the area for EELS mapping scan. Mg capping layer can only be identified with a blurry background due to the low atomic number; c) Cross-sectional HAADF-STEM image of thick-Mg/Ta surface region. *inset*: Zoom-out HAADF-STEM image of thick-Mg/Ta thin film. The green box defines the region of interest for EELS mapping; d) EELS mapping images of thick-Mg/Ta surface region.

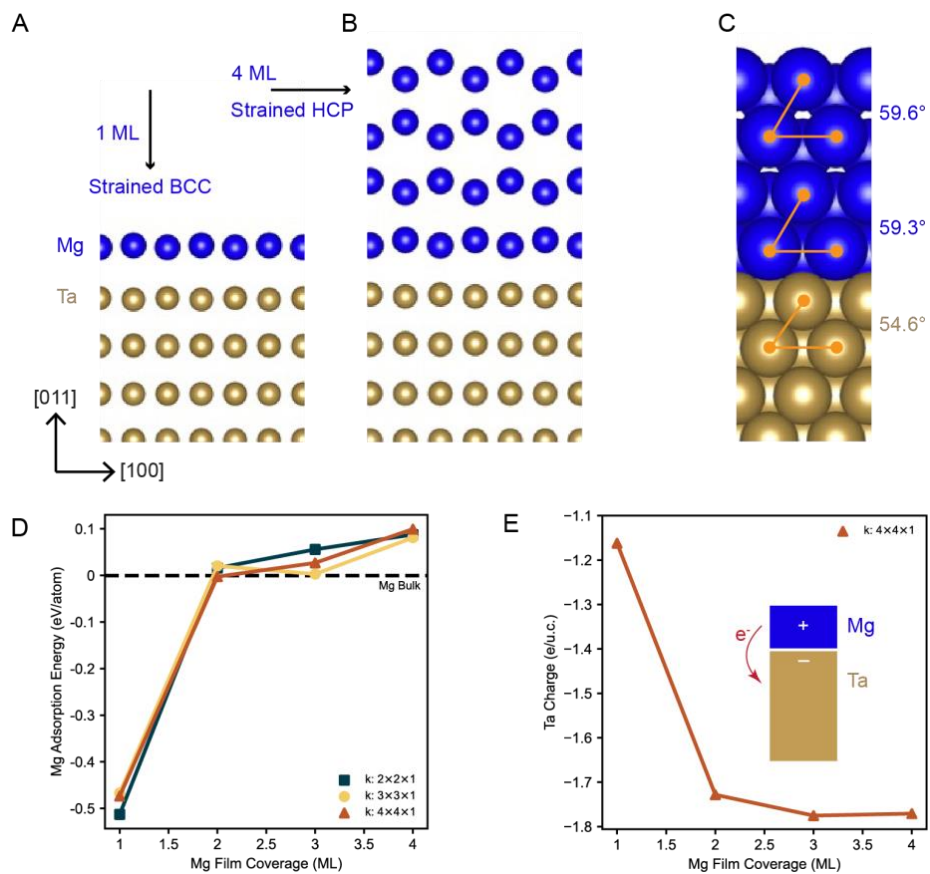

Figure 3. Atomic structures of the Mg/ Ta (110) surface with A) 1 ML and B) 4 ML Mg coverages; C) Angles between Ta-Ta and Mg-Mg bonds indicate gradual transformation of the Mg film to the HCP lattice; D) Adsorption energies (per Mg atom) calculated for the addition of each Mg atomic plane. It is noted that nearly identical values are obtained using 2×2×1, 3×3×1, and 4×4×1 k-meshes; E) Ta charge per lateral Ta (110) unit cell as a function of Mg film coverage.

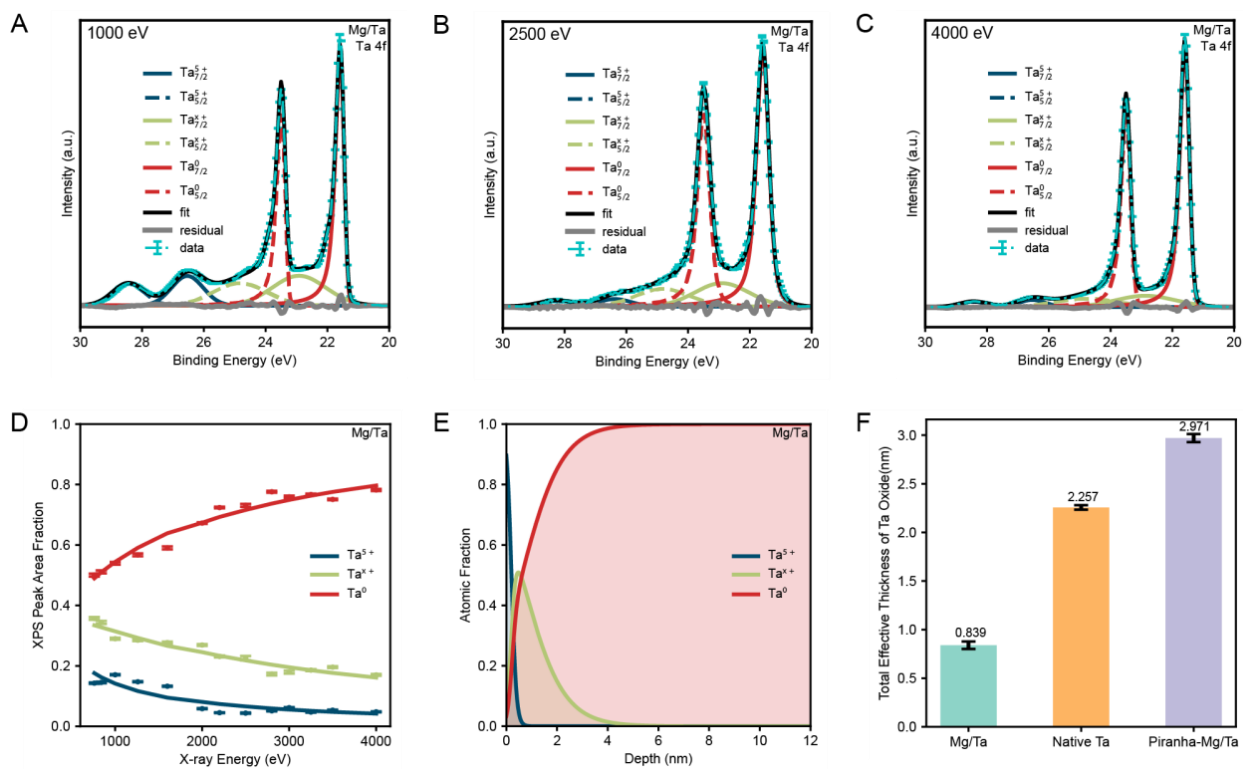

Figure 4. Ta 4f core-level XPS spectra of the Mg/Ta thin film collected at incident X-ray photon energies of A) 1000 eV; B) 2500 eV; and C) 4000 eV. Each spectrum is Shirley background-corrected and fitted with six components, including three Ta species [$Ta^{5+}$ (fully oxidized Ta), $Ta^{x+}$ (Ta suboxides), and $Ta^0$ (metallic Ta)]. For details of background correction and peak fitting, please refer to Ref. [26]. Full dataset is plotted in Figure S6; D) The fitted XPS peak area fractions of different tantalum species in the Mg/Ta thin film as a function of incident X-ray energy. The XPS peak fitting results and corresponding uncertainties are shown in solid dot and error bar, respectively. By modelling the dependence on incident X-ray energy, the peak area fractions are simulated, as shown in solid lines; E) The simulated atomic fractions of different tantalum species as a function of depth of the Mg/Ta thin film. The zero point of depth is at Mg/Ta interface; F) Total effective thickness of tantalum oxide species in the Mg/Ta thin film, Piranha-treated Mg/Ta thin film, and Native Ta thin film. The data of the Native Ta thin film is adapted from Ref. [26]. Note

that the effective thickness is obtained through depth integration, so its value is typically smaller than the x-axis span covered by each corresponding fraction curve in Figure 4E.

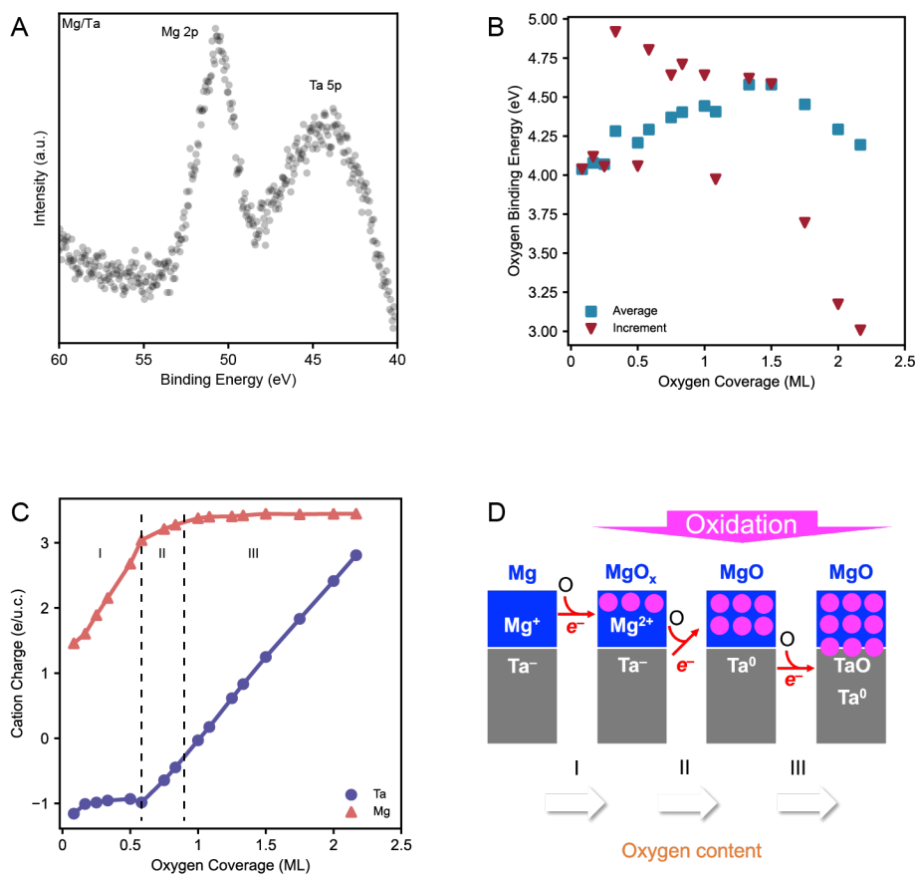

Figure 5. A) Mg 2p core-level XPS spectrum of the Mg/Ta thin film; B) Incremental and average binding energies ($E_i$ and $E_a$, respectively) of oxygen atoms for the case of 1 ML Mg film; $E_i$ values correspond to an additional O at the corresponding oxygen coverage; $E_a$ values are obtained by averaging over binding energies of all adsorbed and absorbed O starting from the pure Mg and up to the given O coverage; C) Atomic charges associated with the Ta and Mg atoms as a function of oxygen content; D) Schematic of the stages of oxidation process. Red arrows indicate the dominant character of the electron density redistribution for each oxidation stage.

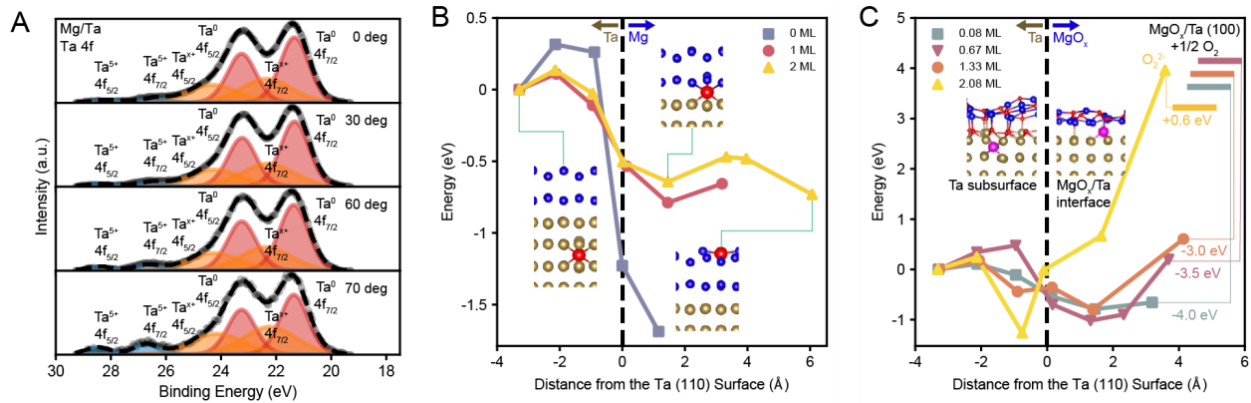

Figure 6. A) Ta 4f core-level angle-resolved XPS spectra of Mg/Ta thin film aged in air for a month at tilting angles of 0°, 30°, 60° and 70°; B) Potential energy profiles for O diffusion from the MgO$_x$ surface into Ta for selected configurations for the low (0.08 ML), medium (0.67 ML and 1.33 ML), and high (2.08 ML) O content. O$_2$ splitting with the formation of O$^{2-}$ ions in MgO$_x$ is favorable for O coverage < 2 ML. The formation of MgO/TaO$_x$ layer (2.08 ML) suppresses O$_2$ dissociation, thus preventing further oxidation; C) Potential energy profiles for out-diffusion of residual oxygen from the Ta layer to the surface (0 Mg ML) and into Mg film (1-2 Mg ML). Insets show the local atomic configurations of the O trapped in Ta (left), O at the Mg/Ta interface (middle), and O at the outer Mg surface (right).

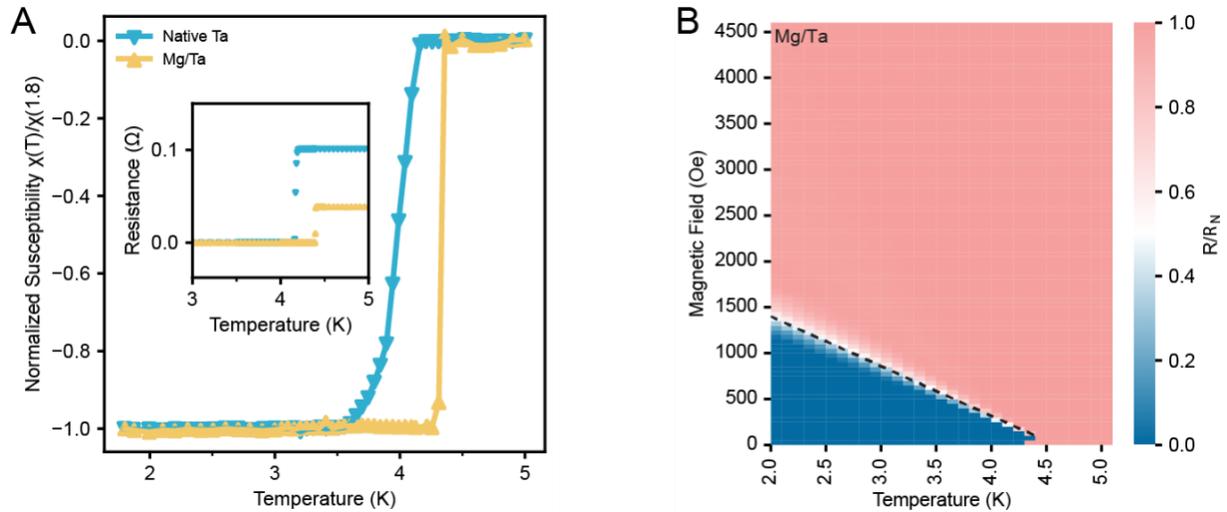

Figure 7. A) The measured thin film susceptibility and resistance (inset) of Native Ta and Mg/Ta as a function of temperature. The susceptibility value at 1.8 K is normalized to unity. B) The dependence of the relative film resistance (normalized by the resistance before superconducting transition at 5 K) of Mg/Ta on magnetic field and temperature. The normal state (pink region) and superconducting state (blue region) are separated by a thin white region denoting the upper critical field ($H_{c2}$). The dotted line represents linear fitting of $H_{c2}$, whose slope is used to calculate the upper critical field at 0 K [$H_{c2}(0)$].

# Supplemental Information for

**Ultrathin Magnesium-based Coating as an Efficient Oxygen Barrier for Superconducting Circuit Materials**

**Materials and Methods**

*Thin film materials synthesis and processing*

All tantalum (Ta) thin film samples used in this study are deposited by magnetron sputtering (AJA Orion), on 2-inch A-plane-cut sapphire wafers (CrysTec, single polished) that are heated to 750 °C. The base pressure of deposition chamber is maintained at approximately $10^{-6}$ Pa. Before loading into the deposition chamber, the sapphire wafers are cleaned by piranha solution prepared by mixing commercial hydrogen oxide solution (30%) and sulfuric acid (98%) in a volume ratio of 1:2, then fully rinsed by deionized (DI) water. After the Ta deposition, the substrate holder is cooled down to approximately 100 °C, followed by sputtering deposition of the magnesium (Mg) capping layer. Ta and Mg sputtering targets are obtained from AJA International, both having purities of 99.98%. Oxidation resistance of the Mg/Ta film is tested by either leaving the film in room temperature atmosphere for 30 days or heating to 200 °C for 5 min in an atmospheric box furnace. To test the resistance to acid processing, the thin films are fully immersed in piranha solution, then rinsed by deionized water and dried by nitrogen gas sufficiently.

*Materials Characterizations*

X-ray diffraction (XRD) patterns of the thin films are collected on a Rigaku SmartLab II X-ray diffractometer using Cu Kα radiation ($\lambda = 1.5418$ Å). Transmission electron microscopy (TEM) and electron energy loss spectroscopy (EELS) studies are carried out on a JEOL ARM-200F, with the specimen prepared by the focused ion beam (FIB) lift-off technique using a dual beam SEM/FIB microscope (FEI Helios). The thickness (*t*) of the thin film and the mapping region are measured using ImageJ. Each measurement consists of averaging ten data points and calculating the corresponding standard deviation following the equations:

$$\bar{t} = \frac{\sum_{i=0}^{10} t}{10}, \tag{1}$$

$$\sigma_t = \sqrt{\frac{\sum_i^{10}(t-\bar{t})^2}{9}}. \tag{2}$$

All lab-based X-ray photoemission spectroscopy (XPS) measurements, including angle-resolved XPS (AR-XPS) and ambient-pressure XPS (AP-XPS), use Al Kα ($hv$ = 1486.6 eV) as the excitation source. AR-XPS is performed in a RHK multiprobe system equipped with a tilt stage (Figure S10a). The initial angle between the X-ray source and the sample stage is 45°. The X-ray incident angle ($\varphi$) and the photoelectron collection angle ($\alpha$) are both geometrically related to the sample state tilting angle ($\theta$), following the equations:

$$\varphi = 90° - \arccos(\cos\theta/\sqrt{2}), \tag{3}$$

$$\alpha = 90° - \theta. \tag{4}$$

According to the quantitative relationship, the angle of incident X-ray and the photoelectron collection angle both monotonically decrease as the sample stage tilts around an axis in the X-ray plane (Figure S10b). As such, the surface sensitivity is enhanced as the sample tilt angle increases because the effective information depth for both incident photons and photoelectrons are both shallower at smaller grazing angles of incidence and emission. AP-XPS is performed in a UHV chamber, with base pressure kept below $1 \times 10^{-6}$ Pa. The spectrometer is equipped with a differentially pumped hemispherical analyzer (SPECS PHOIBOS NAP 150). Lab-based XPS data is analyzed using software CasaXPS Version 2.3.23. All spectra are calibrated with gold reference by setting the binding energy of Au $4f_{7/2}$ peak to 84.0 eV and fitted with Shirley background.

Variable-energy XPS (VEXPS) is performed on the Spectroscopy Soft and Tender (SST) beamline operated by the National Institute of Standards and Technology (NIST) at National Synchrotron Light Source II (NSLS-II), Brookhaven National Laboratory (BNL). The measurements are conducted at the hard X-ray photoelectron spectroscopy (HAXPES) endstation, using X-ray beams with energy ranging from soft regime (photon energy < 2 keV, delivered by SST-1 beamline) to tender regime (photon energy ⩾ 2 keV, delivered by SST-2 beamline). At the HAXPES endstation, takeoff angle is fixed at 80°, regardless of X-ray beam source. The incident angle of X-ray beam is 6° for SST-1 and 10° for SST-2. In the XPS trace analysis, we first calibrate the binding energy of each spectrum with gold reference by setting the binding energy of Au $4f_{7/2}$ peak to 84.0 eV, then subtract a Shirley background. Each Ta 4f spectrum is fitted with three components, including $Ta^{5+}$ (fully oxidized Ta), $Ta^{x+}$ (Ta suboxides), and $Ta^0$ (metallic Ta). The oxidized Ta species are modelled with Gaussian profiles, while the metallic Ta is modelled with skewed Voigt profile. Each O 1s spectrum is fitted with two components, including lattice $O^{2-}$ (O) and hydroxyl group (OH), both modelled with Gaussian profiles. The depth profile is simulated using a double-attenuation model that involves incident photon attenuation and photoelectron attenuation. For more details of peak fitting, uncertainty estimation and depth profiling, please refer to the work Ref. 1.

Low-temperature electrical transport measurements are carried out with Physical Property Measurement System (PPMS, Dynacool™ from Quantum Design). The dc resistance is measured using the DC Resistivity Option. The transition temperature, $T_c$, is determined by the criteria of superconducting transition midpoint, $R/R_N=0.5$, where $R_N$ is the resistance before superconducting transition at 5 K. The residual-resistance ratio (RRR) is determined by the ratio between the resistance at room temperature and $R_N$. Low-temperature magnetic properties are characterized by Magnetic Property Measurement System (MPMS, equipped with SQUID Magnetometer from Quantum Design).

The critical field is estimated using Ginzburg-Landau theory. A type-II superconductor is typically characterized with a shorter effective coherence length ($\xi$), which is related to the Pippard coherence length ($\xi_0$) and the electron mean free path ($l$) through

$$\xi(T) = 0.855\sqrt{\xi_0 l/(1 - T/T_c)}, \tag{5}$$

in the dirty limit.[2] As impurity level increases, electron mean free path $l$ will decrease, followed by the effective coherence length $\xi$, which eventually turns a bulk type-I superconductor to type-II.[3] Thus, the dc magnetic susceptibility again demonstrates that Mg capping lowers the impurity content in the Ta film. For a type-II superconductor, the effective coherence length at 0 K, $\xi(0)$, can be estimated by the Ginzburg-Landau expression,

$$\xi(0) = \sqrt{\frac{\Phi_0}{2\pi H_{c2}(0)}}, \tag{6}$$

where $\Phi_0 = h/2e$ is the magnetic flux quantum and $H_{c2}(0)$ is the upper critical field at 0 K. The latter can be extracted from the dependence of the film resistance ($R$) on magnetic field ($H$) and temperature ($T$), respectively shown in Figure 6b and Figure S14a for Mg/Ta and Native Ta, using the Werthamer-Helfand-Hohenberg model:

$$H_{c2}(0) = -\frac{\pi T_c(0)}{8e^\gamma} \frac{dH_{c2}}{dT}\bigg|_{T=T_c(0)}, \tag{7}$$

in which $T_c(0)$ is the zero field $T_c$, $\gamma$ is the Euler's constant (0.5572), and the derivative term is determined by linear fit to the $H_{c2} - T$ relation near $T = T_c(0)$. The analysis finds $H_{c2}(0)$ at 0.16 T and 0.25 T, for Mg/Ta and Native Ta, respectively. Following Eq. 6, we further determine $\xi(0) = $ 45 nm for Mg/Ta and 36 nm for Native Ta. Given that the 0 K London penetration depth, $\lambda_L(0)$, is 35 nm for Ta[4], the Ginzburg-Landau (GL) parameter, $\kappa = \lambda_L(0)/\xi(0)$, is found at 0.78 for Mg/Ta and 0.97 for Native Ta. Since both values are larger than $1/\sqrt{2} \approx 0.707$, we may confirm that Mg/Ta and Native Ta are both type-II semiconductors. However, the GL parameter of Mg/Ta is very close to the $1/\sqrt{2}$ threshold, which may explain its sharp transition to the Meissner state in the magnetic susceptibility measurement (Figure 6a). The electron mean free path, on the other hand, is estimated at 36 nm and 21 nm, respectively for Mg/Ta and Native Ta, using Eq. 7 and the Pippard coherence length of bulk Ta ($\xi_0$ = 78 nm)[4]. This further confirms the previous assessment that the superfluid transport in Native Ta is more affected by impurities.

*DFT Calculations*

The pure Mg/Ta (110) and oxidized MgO$_x$/Ta (110) structures are represented using the periodic slab model. The Ta slab is four atomic planes thick; the lateral supercell corresponds to the 2×3 extension of the BCC (110) primitive cell and contains 12 Ta atoms per plane. The out-of-place supercell parameters is 2.5 nm, leaving approximately 1.3 nm vacuum gap. Stages of the oxidation process are simulated by gradually increasing oxygen content at the surface and in the subsurface region. Ta atoms of the bottom plane of the slab are fixed in their bulk atomic positions, while the coordinates of all other atoms are fully relaxed. The calculations are performed using the VASP[5] package and Perdew–Burke-Ernzerhof (PBE)[6] exchange correlation functional. The projector-augmented wave potentials were used to approximate the effect of the core electrons.[7] We test convergence of the calculated adsorption energies with respect to the density of k-mesh for Mg binding to Ta (110); energies calculated using 2×2×1, 3×3×1, and 4×4×1 k-meshes are nearly identical. Therefore, the 2×2×1 k-mesh is used in all subsequent calculations. The plane-wave basis set cutoff is set to 500 eV and the total energy convergence criterion is set to $10^{-5}$ eV. Bader charge population analysis is used to analyze charge density redistribution.[8] The oxygen binding energies are calculated with respect to the triplet state of the gas-phase O$_2$ molecule.

**Disclaimer**

Commercial equipment, instruments, or materials are identified in this paper to specify the experimental procedure adequately. Such identification is not intended to imply recommendation or endorsement by the National Institute of Standards and Technology, nor is it intended to imply that the materials or equipment identified are necessarily the best available for the purpose.

**Figures**

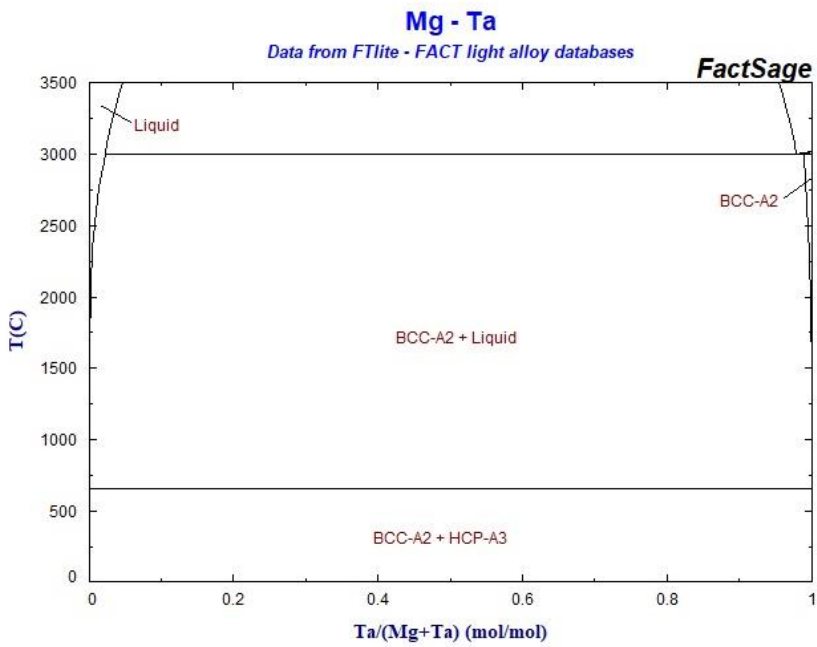

Figure S1. Ta/Mg phase diagram adapted from Facility for the Analysis of Chemical Thermodynamics (FACT) light alloy database.[9]

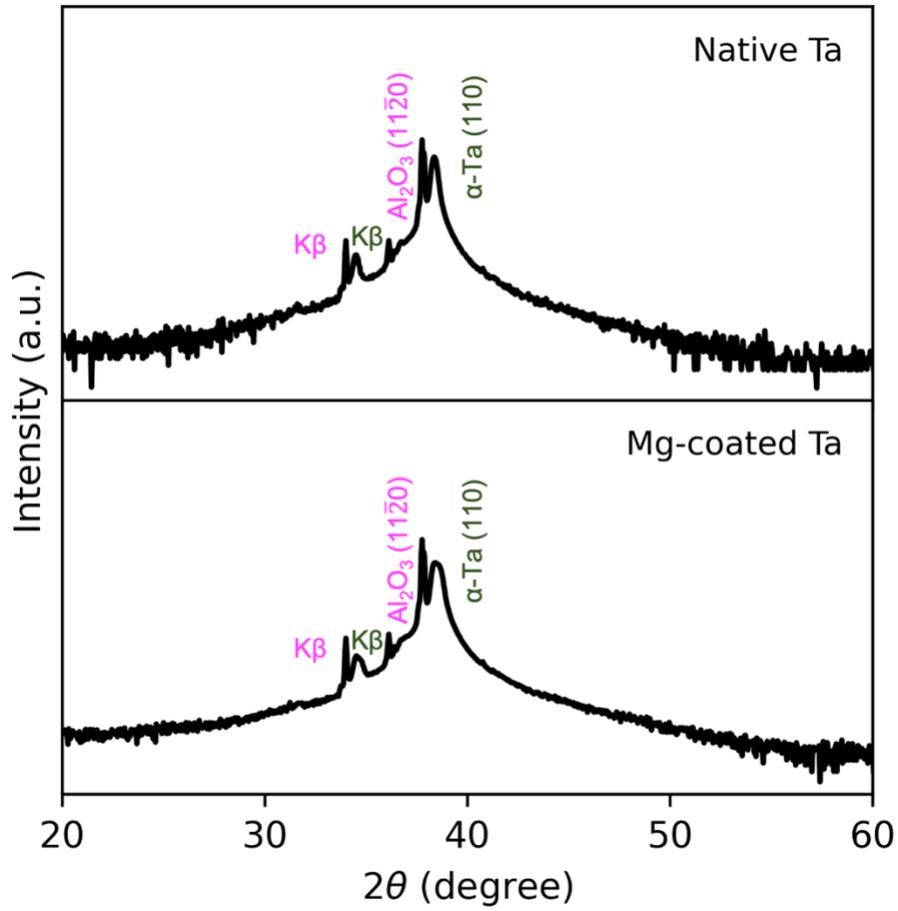

Figure S2. XRD patterns of Native Ta (top) and Mg/Ta (bottom) thin film.

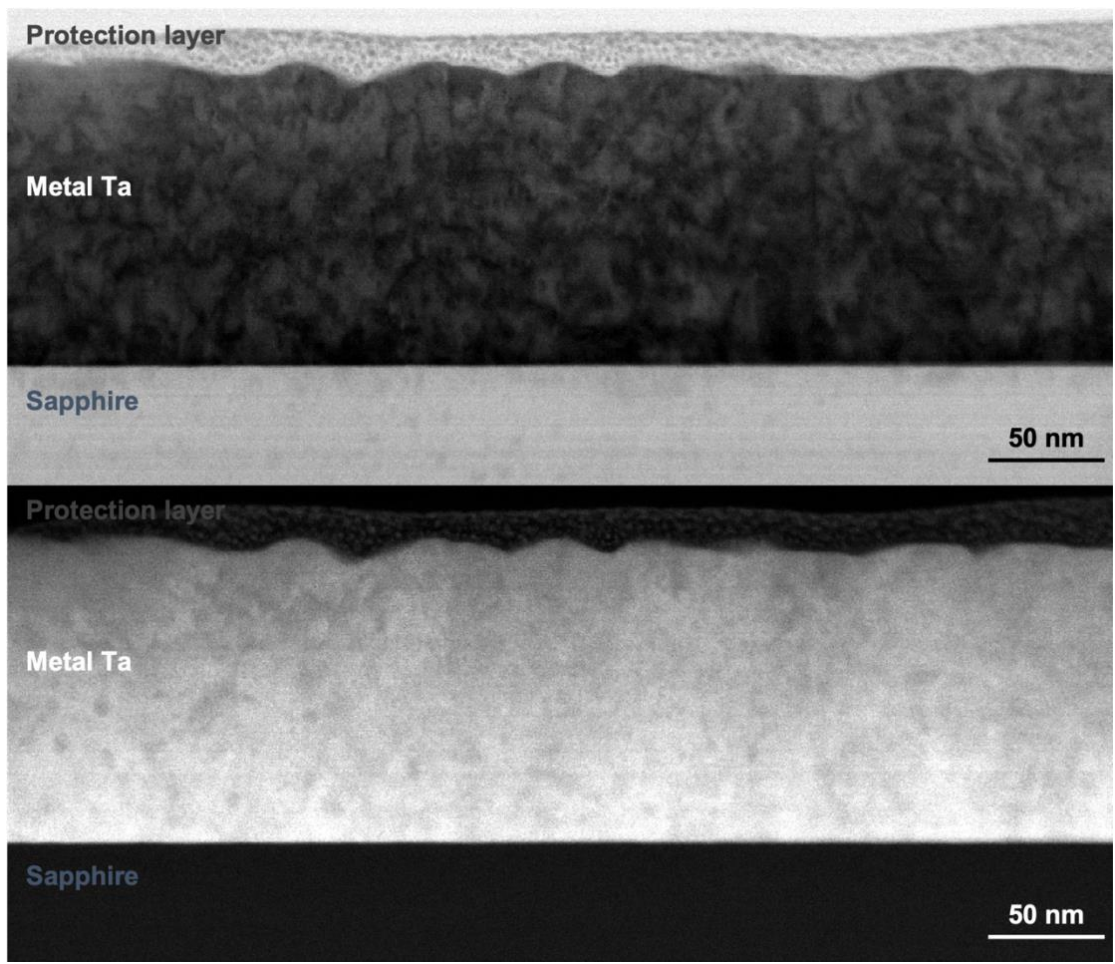

Figure S3. Cross-sectional ABF (top) and HAADF (bottom) STEM images of Mg/Ta thin film.

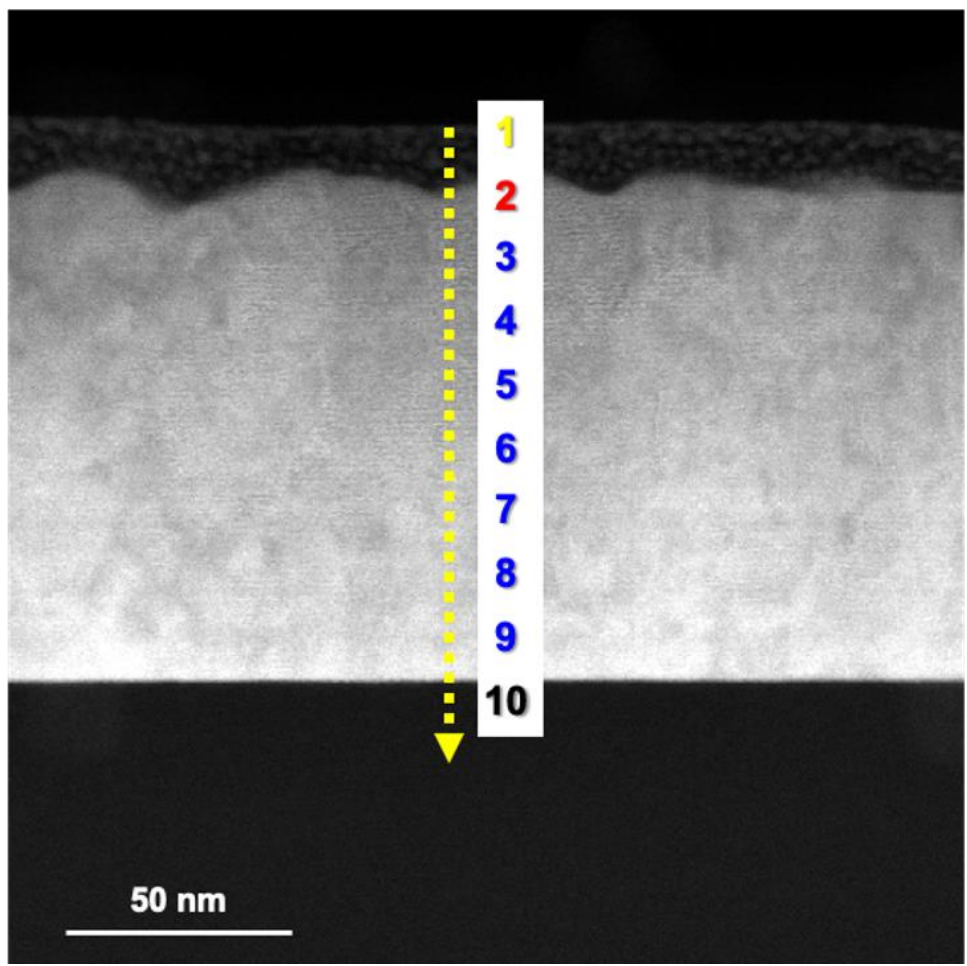

Figure S4. Cross-sectional HAADF-STEM image of Mg/Ta thin film. The yellow arrow indicates the scanning line and direction for EELS line scan. The colored numbers show the location for each scan (#1-#10 from top to bottom).

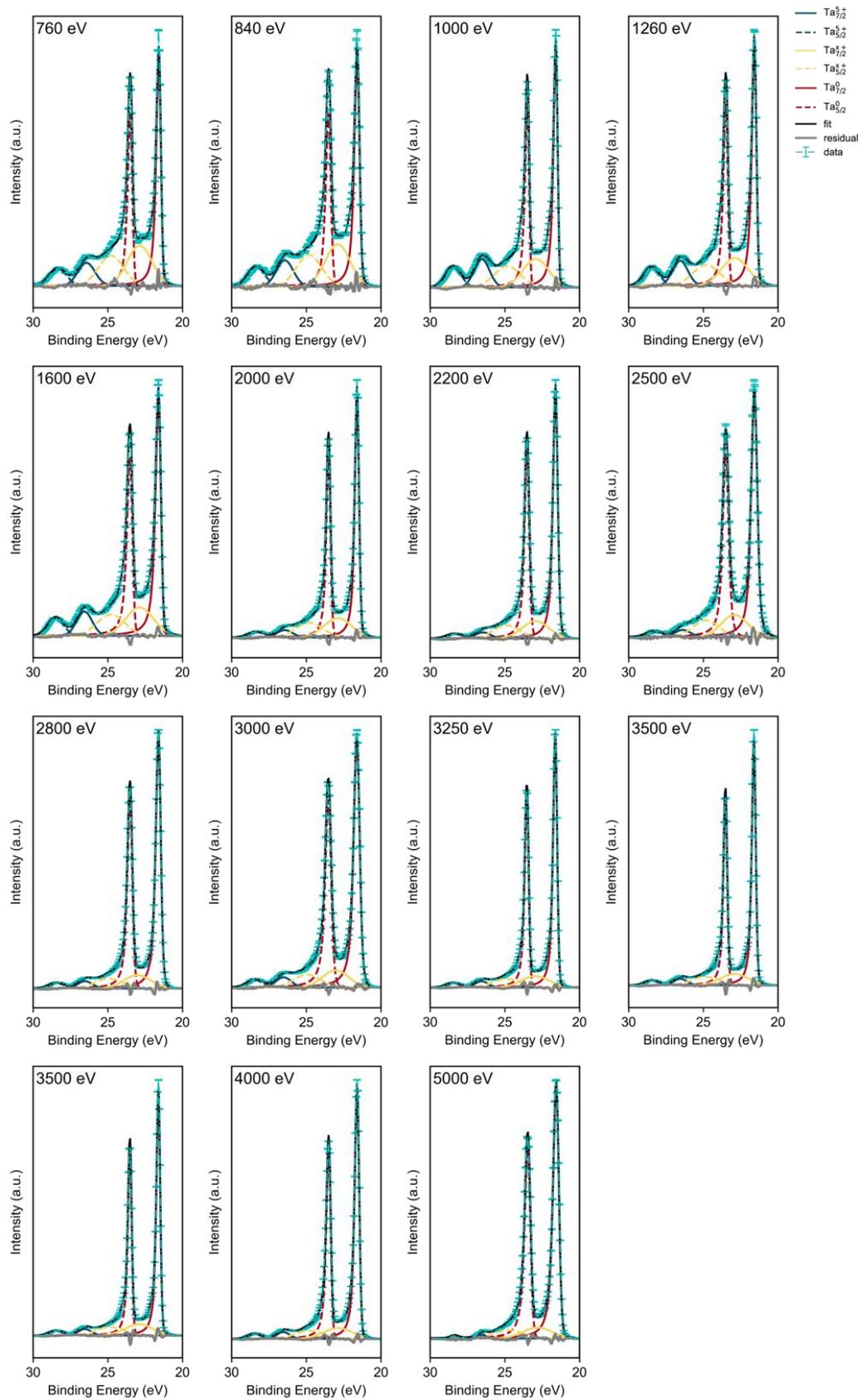

Figure S5. Full dataset of Ta 4f core-level XPS spectra collected at 15 incident X-ray photon energies. Each spectrum is Shirley background-corrected and fitted with five components, including $Ta^{5+}$, $Ta^{x+}$, and pure metal Ta ($Ta^0$).

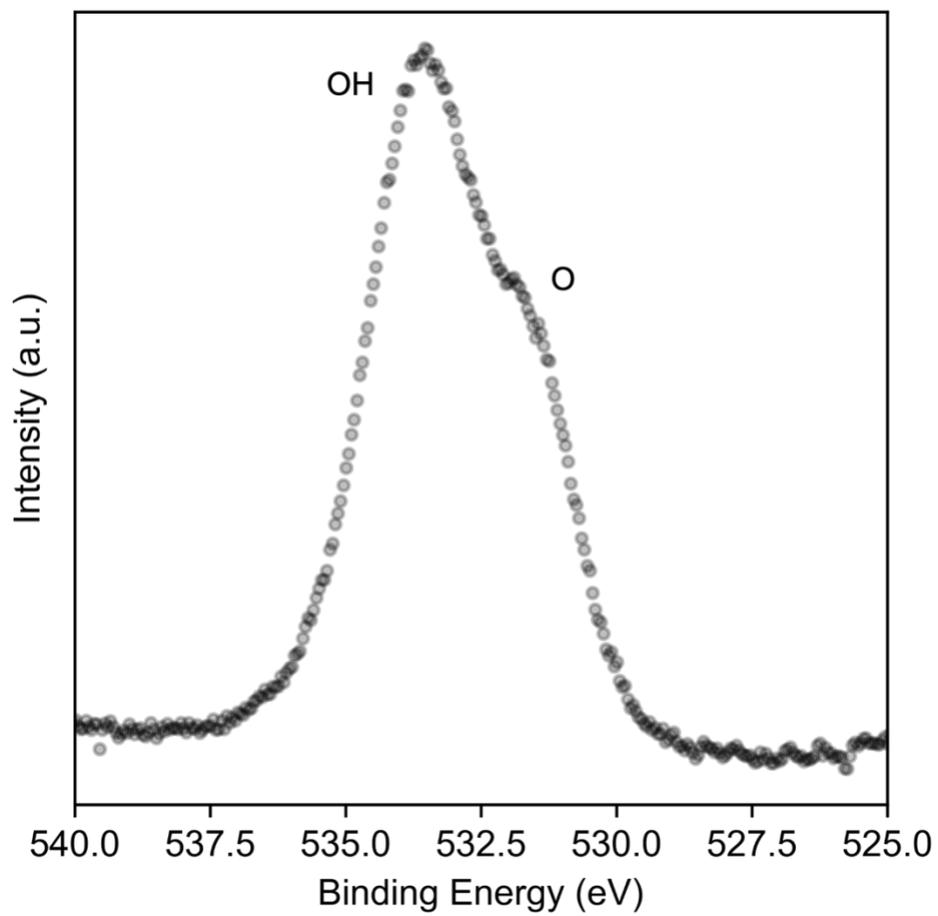

Figure S6. O 1s core-level XPS spectrum of Mg/Ta thin film.

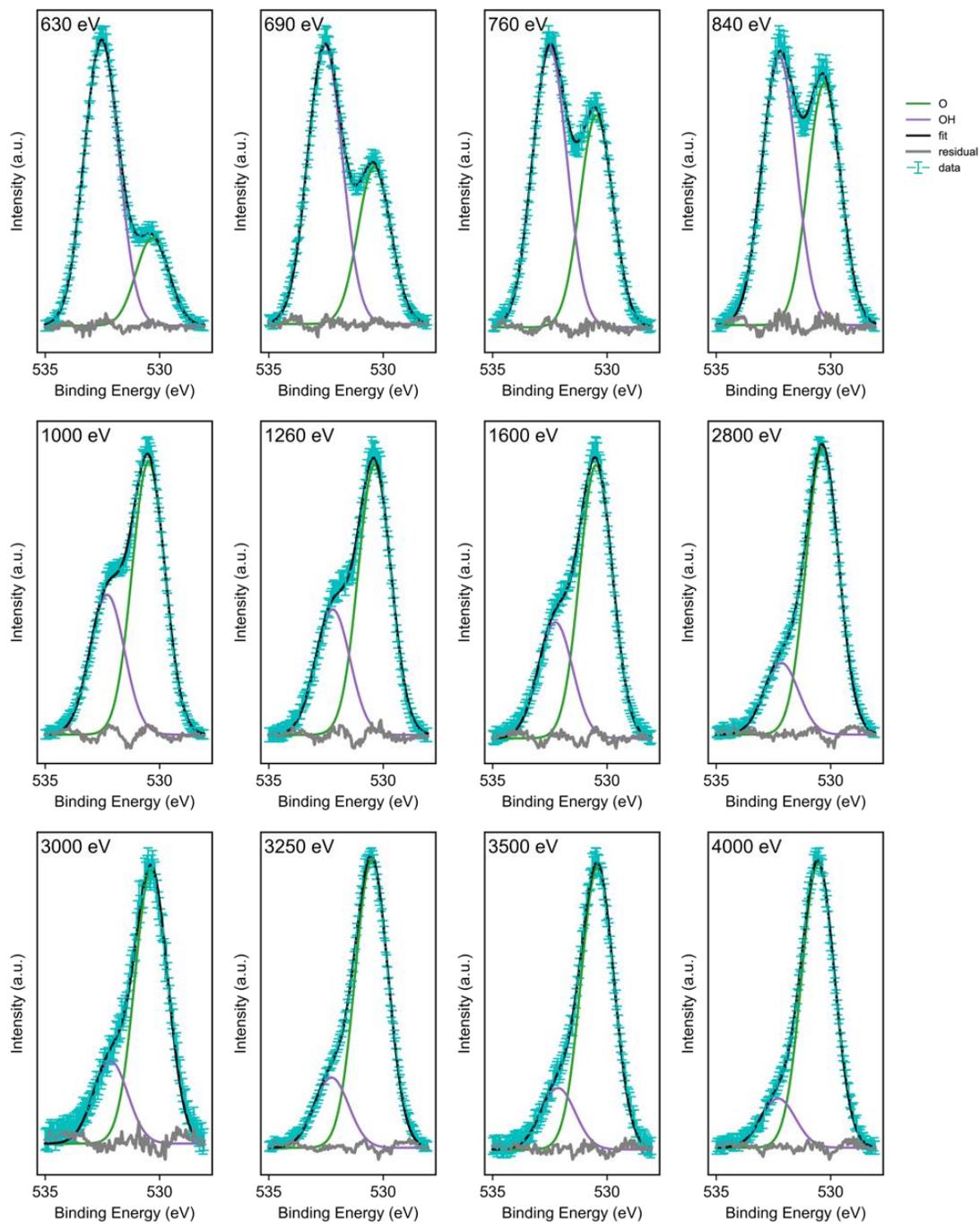

Figure S7. Full dataset of O 1s core-level XPS spectra collected at 12 incident X-ray photon energies. Each spectrum is Shirley background-corrected and fitted with two components, including lattice $O^{2-}$ (O) and hydroxyl group (OH).

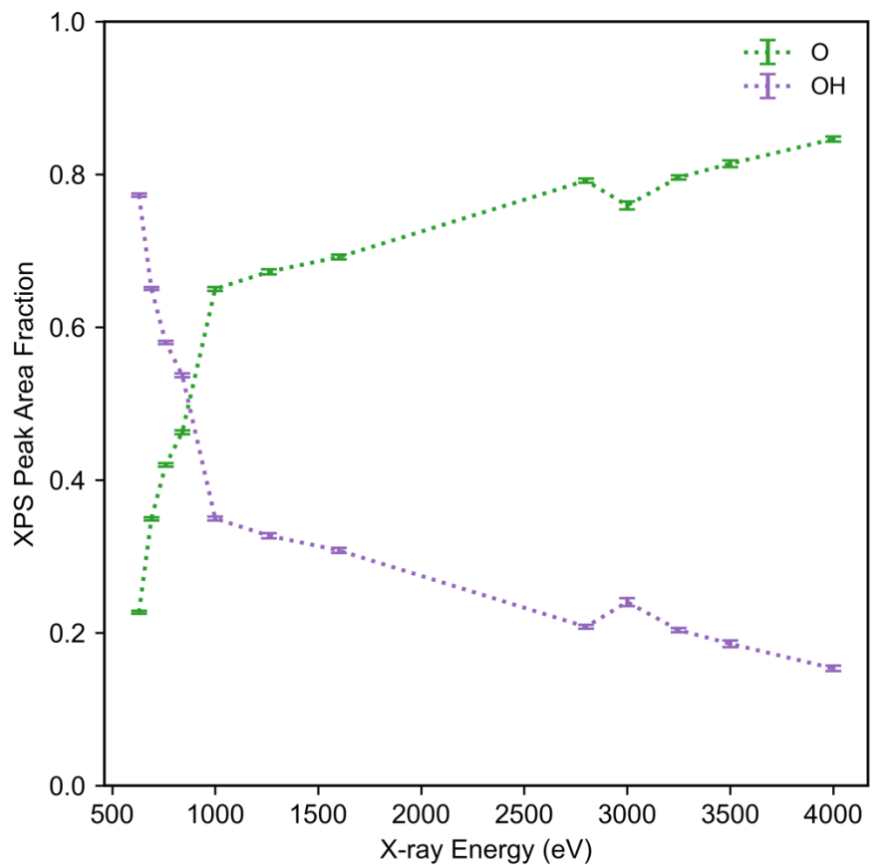

Figure S8. The fitted XPS peak area fractions of different oxygen species as a function of incident X-ray energy of Mg/Ta thin film. The XPS peak fitting results and corresponding uncertainties are shown with solid dot and error bar, respectively.

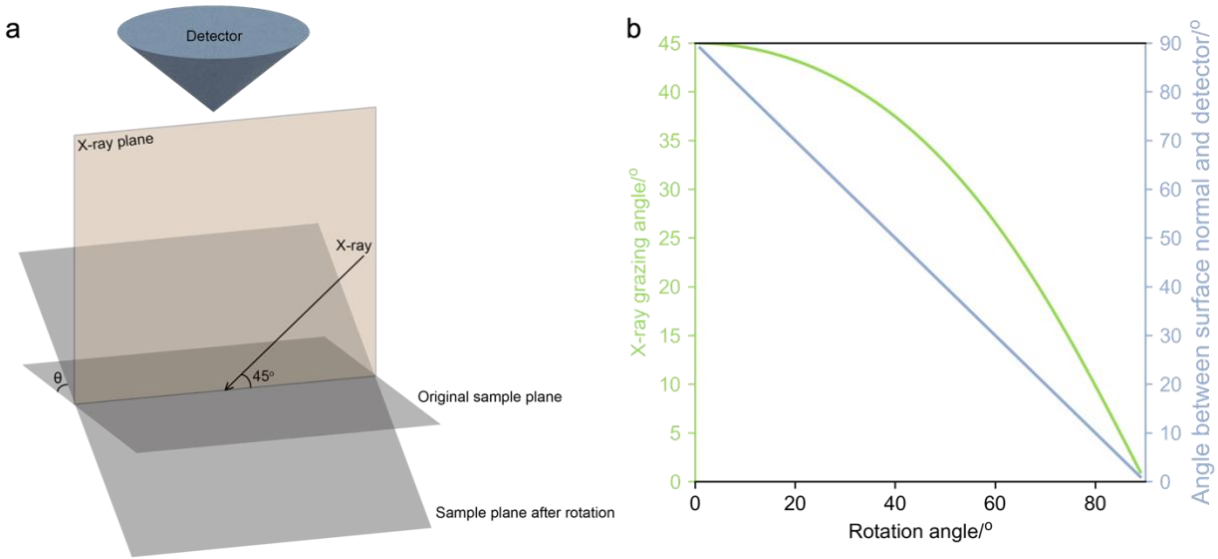

Figure S9. a) Schematic of angle-resolved XPS (AR-XPS). In the original layout, the sample is horizontal with X-ray illuminating at 45°, and the detector is at the normal of sample stage. The axis for sample tilting, whose degree is denoted by θ, lies in the X-ray plane; b) The dependence of X-ray grazing angle and the angle that between surface normal and detector on the sample tlting angle. Both angles are monotonically decreasing as the sample stage tilts.

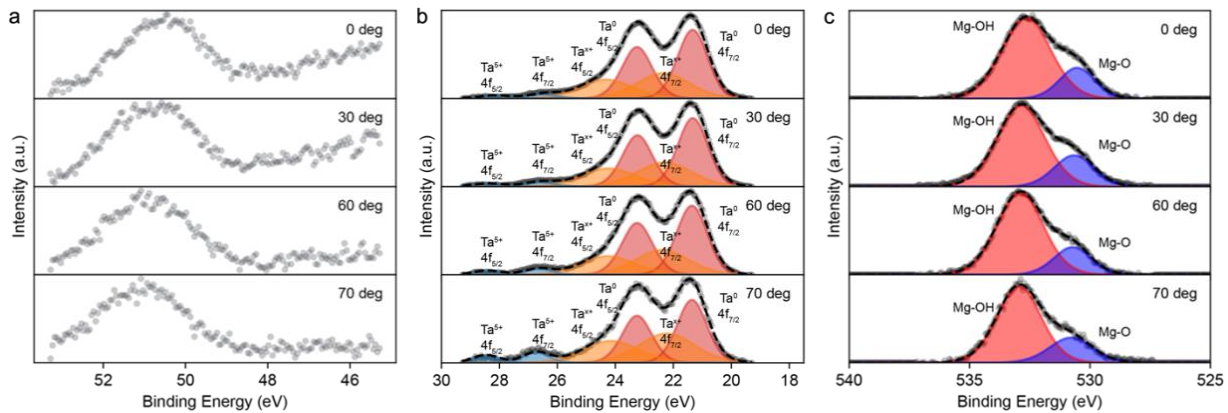

Figure S10. a) Mg 2p, b) Ta 4f and c) O 1s core-level VAXPS spectra of Mg/Ta thin film aged in air for a month at tilting angles of 0°, 30°, 60° and 70°.

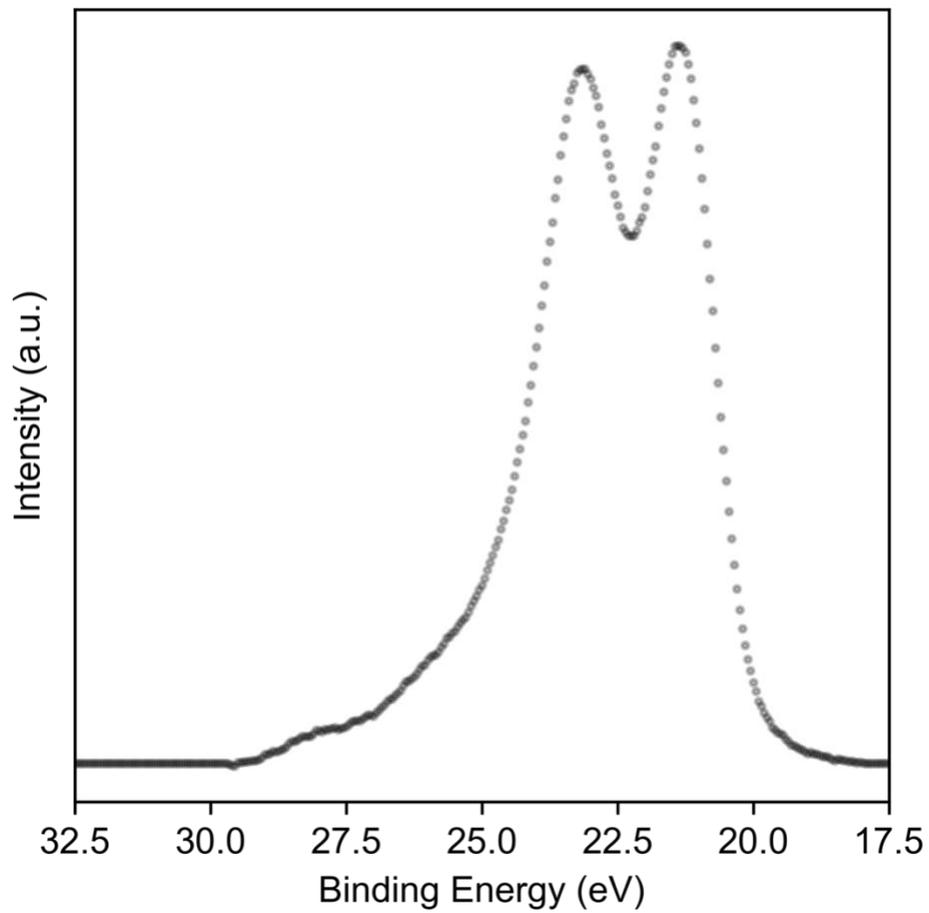

Figure S11. Ta 4f core-level XPS spectra of Mg/Ta thin film annealed in air for 5 minutes at 200 °C.

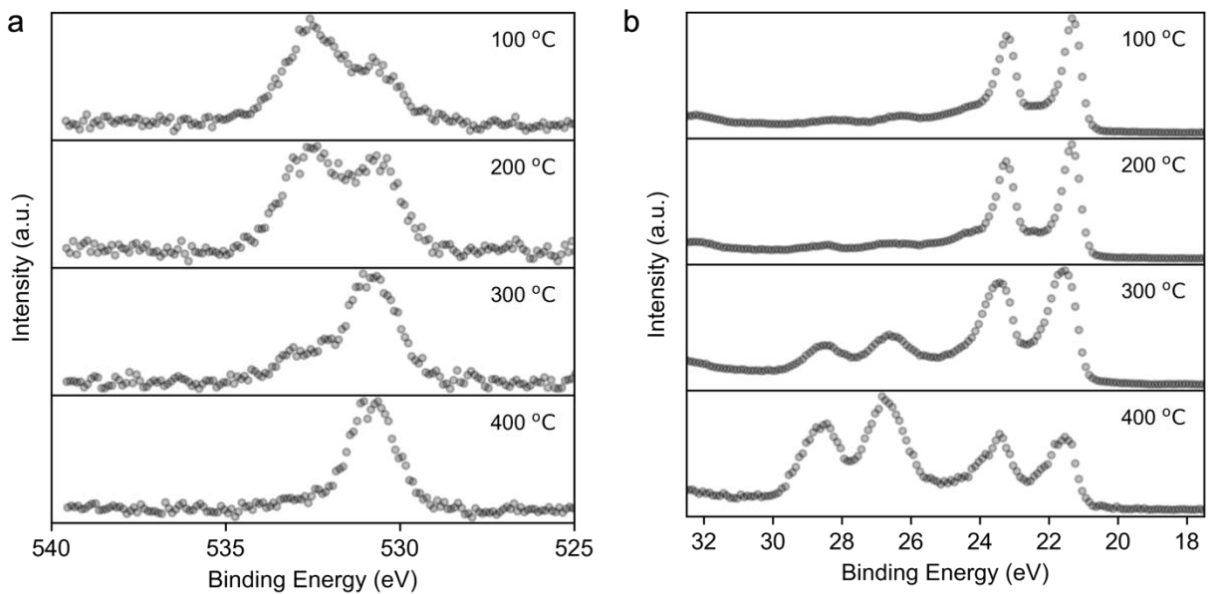

Figure S12. a) O 1s core-level APXPS spectra of Mg/Ta thin film under the $O_2$ pressure of 50 Pa at 100 °C, 200 °C, 300 °C and 400 °C; b) Ta 4f core-level APXPS spectra of Mg/Ta thin film under the $O_2$ pressure of 50 Pa at 100 °C, 200 °C, 300 °C and 400 °C.

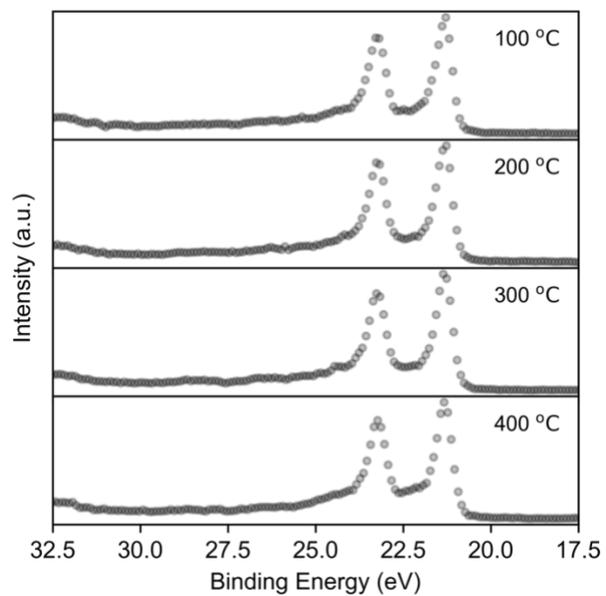

Figure S13. Ta 4f core-level APXPS spectra of Mg/Ta thin film under ultra-high vacuum at 100 °C, 200 °C, 300 °C and 400 °C.

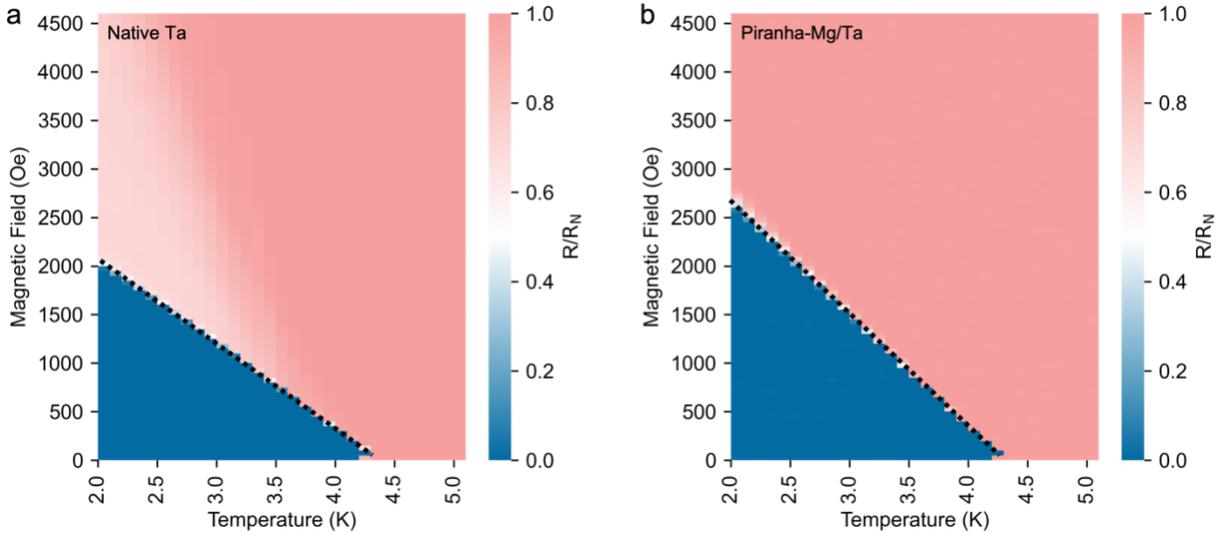

Figure S14. The dependence of the relative film resistance (normalized by the resistance before superconducting transition at 5 K) of a) Native Ta and b) Piranha-Mg/Ta on magnetic field and temperature.

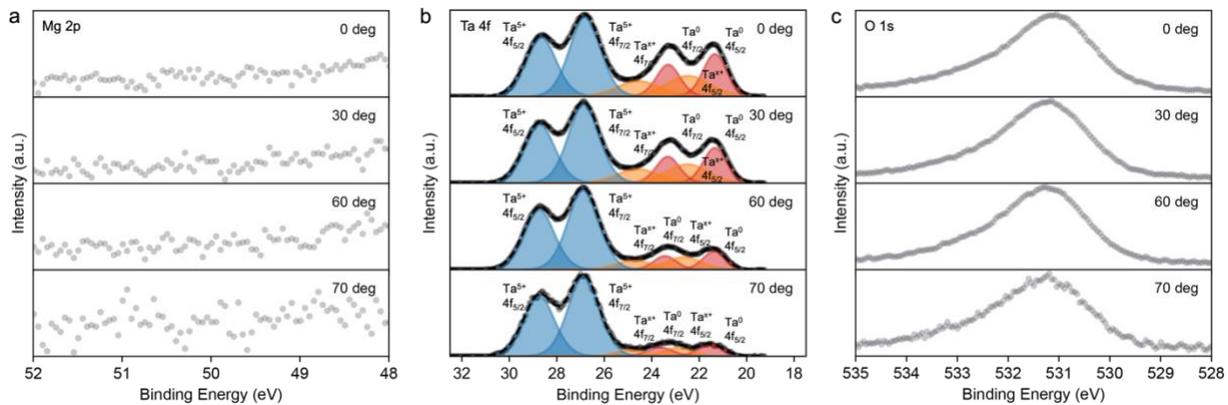

Figure S15. a) Mg 2p, b) Ta 4f and c) O 1s core-level VAXPS spectra of Mg/Ta thin film after piranha treatment at tilting angles of 0°, 30°, 60° and 70°.

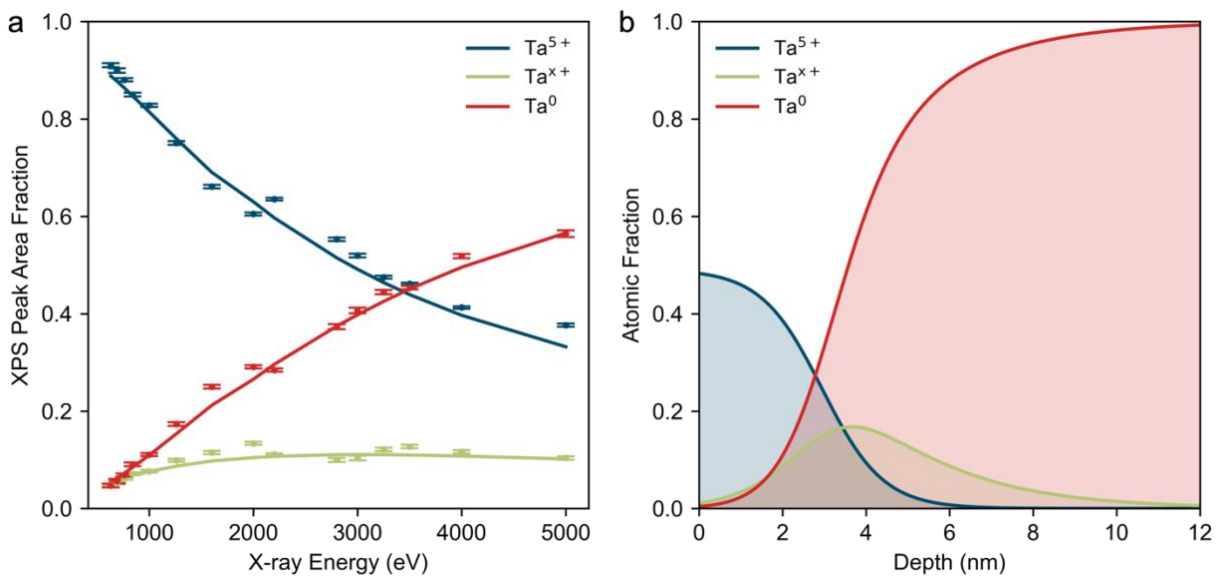

Figure S16. a) The fitted XPS peak area fractions of different tantalum species as a function of incident X-ray energy of the Piranha-Mg/Ta thin film. The XPS peak fitting results and corresponding uncertainties are shown in solid dot and error bar, respectively. By modelling the dependence on incident X-ray energy, the peak area fractions are simulated, as shown in solid lines; b) The simulated atomic fractions of different tantalum species as a function of depth of the Piranha-Mg/Ta thin film. The zero point of depth is at Mg/Ta interface. The values of effective thickness estimated from fitting are shown in Figure 4f.

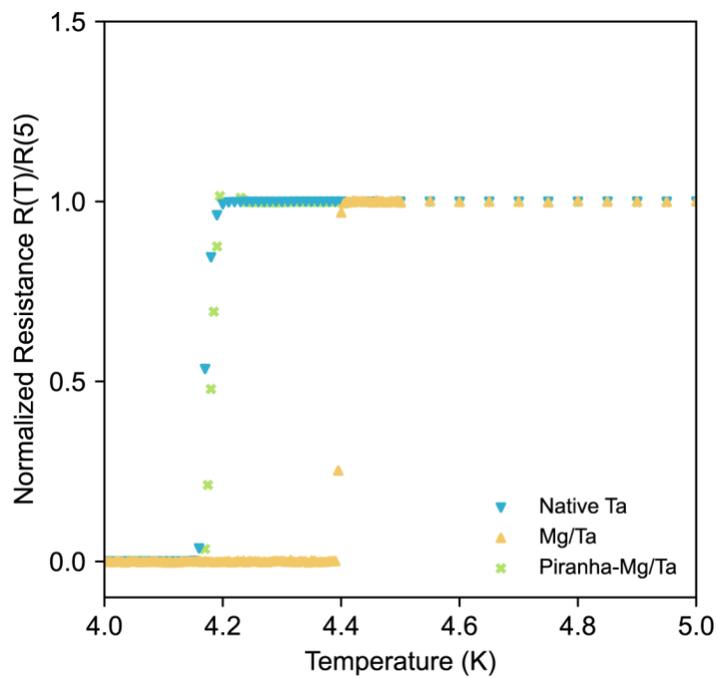

Figure S17. Normalized thin film resistance of Native Ta, Mg/Ta and Piranha-Mg/Ta as a function of temperature.